\documentclass{PoS}
\usepackage{graphicx} 
\usepackage{epsfig}
\usepackage{subfig}

\title{The GAPS Project: First Results}

\ShortTitle{The GAPS Project}

\author{{\speaker{S. Benatti}$^1$\thanks{SB acknowledges the support from INAF trough the "Progetti Premiali" funding scheme of the
Italian Ministry of Education, University, and Research.} , R.Claudi$^1$, S. Desidera$^1$, R. Gratton$^1$, A.F. Lanza$^2$, G. Micela$^3$, I. Pagano$^2$, G. Piotto$^4$, A. Sozzetti$^5$, C. Boccato$^1$, R. Cosentino$^6$, E. Covino$^7$, A. Maggio$^3$, E. Molinari$^6$, E. Poretti$^8$, R. Smareglia$^9$ and the GAPS Team} \\

      $^1$  INAF - Astronomical Observatory of Padova\\
      $^2$ INAF - Astrophysical Observatory of Catania\\
      $^3$  INAF - Astronomical Observatory of Palermo\\
      $^4$  University of Padova, Dep. of Physics and Astronomy \\
      $^5$  INAF - Astronomical Observatory of Torino\\
      $^6$  INAF - TNG, Fundaci\'on Galileo Galilei \\
      $^7$  INAF - Astronomical Observatory of Capodimonte\\
      $^8$  INAF - Astronomical Observatory of Brera \\
      $^9$  INAF - Astronomical Observatory of Trieste\\
        E-mail: \email{serena.benatti@oapd.inaf.it}}

\abstract{The GAPS programme is an Italian project aiming to search and characterize extra-solar planetary systems around stars with different characteristics (mass, metallicity, environment). GAPS was born in 2012, when single research groups joined in order to propose a long-term multi-purpose observing program for the exploitation of the extraordinary performances of the HARPS-N spectrograph, mounted at the Telescopio Nazionale Galileo. Now this group is a concerted community in which wide range of expertise and capabilities are shared in order to reach a more important role in the wider international context. We present the results achieved up to now from the GAPS radial velocity survey: they were obtained in both the two main objectives of the project, the planet detection and the characterization of already known exoplanetary systems. With GAPS we detected, for instance, the first confirmed binary system in which both components host planets \cite{desidera}, the first planetary system around a star in an open cluster \cite{malavolta}, a system of Super-Earths orbiting an M-dwarf star \cite{affer}. 

      }

\FullConference{Frontier Research in Astrophysics -- II\\
		23-28 May 2016\\
		Mondello (Palermo), Italy}

\begin{document}

\section{Introduction}

Since 2012, after the installation of the high resolution echelle spectrograph HARPS-N \cite{cosentino}, the Italian telescope TNG (La Palma, Canary Islands) became one of the key facilities for the study of the extrasolar planets. The opportunity to exploit the great capabilities of this instrument encouraged a large fraction of Italian scientists working in the exoplanets field to set-up a collaboration and to propose a long-term observing program with specific themes and objectives. The "Global Architecture of Planetary Systems" (GAPS) Project is the result of this collaboration. It gathers more than 60 astronomers from several institutes of the Italian National Institute for Astrophysics (INAF) and Italian Universities (Padova, Torino and Milano). A technical and scientific support is also provided by a few collaborators from  European and American Institutions. 

The main purpose of GAPS is the study and the characterization of the architectural properties of planetary systems through the radial velocity technique, by analyzing the distributions of planetary parameters and their correlations with those of the host star. The opportunity to extend our knowledge of the planetary systems will help to understand the most debated aspects of the exoplanet research, such as their formation and evolution.

\subsection{Scientific scenario}
The results of the NASA-Kepler mission show that our Solar System is not a model for other exoplanetary systems, but a large variety of planetary sizes and systems architectures actually exists (\cite{howard}, \cite{fabrycky}, \cite{batalha}). In particular, low-mass planets (in the range between Super-Earths and Neptunes) appear to be ubiquitous and are generally found in multiple systems, tightly packed close to their central star \cite{howard2}. The improvements in the technological capabilities, together with the complementarity of the techniques, played a crucial role in the discovery of such a diversity, allowing to push the detection limits toward planets with lower masses and wider orbits.
In this framework the radial velocity (RV) technique, which measures the reflex motion of the star induced by a planet, has provided a large number of discoveries and represents an essential tool for the follow-up of transit candidates.

The radial velocity of the star orbiting the barycenter of the planet - star system changes as a function of time with a period equal to the orbital period of the planet, $P$, and a semiamplitude, $K$, defined as follows:
$$
K = \bigg(\frac{2\pi G}{P}\bigg)^{1/3} \frac{M_{\rm p} \sin i}{(M_{\rm p}+M_{\star})^{2/3}} \frac{1}{(1-e^2)^{1/2}}
$$
$M_{\rm p}$ being the planetary mass (since the real mass depends on the unknown inclination, $i$, of the orbit, the RV technique only provides a value for the minimum mass), $M_{\star}$ the stellar mass and $e$ the orbital eccentricity of the planet. This method is more sensitive to high planetary mass and short orbital period: a Jupiter-like planet orbiting at 1 AU from its host star produces a RV semiamplitude of about 30 m s$^{-1}$, while an Earth-like planet at the same distance induces a RV modulation of only 9 cm s$^{-1}$. 
The detection of multiple systems with low-mass components requires a particular observational effort ($\sim100$ RV epochs) in order to separate the single signals.
Moreover, the RV modulation due to the contribution of the stellar activity can mimic the presence of a planetary companion, hampering the interpretation of the time series.

An accurate observing strategy, focused on an intensive monitoring, the adoption of new sophisticated techniques for proper treatment and mitigation of activity-induced signals, and the use of high resolution spectrographs are the key points to successfully detect and confirm the presence of low mass planetary companions in a multi-planet system.

\section{The contribution of GAPS}
The GAPS observing programme started in September 2012. Since then $\sim2500$ observing hours were allocated and about 7000 spectra have been collected for the targets of our sample ($\sim 300$ stars). For 16 interesting objects we have obtained 90 or more RV data points.
The GAPS observations are performed with HARPS-N, the twin instrument in the northern hemisphere of HARPS at the 3.6m ESO-La Silla telescope. These two spectrographs are fiber-fed and work in the the visible range (400-600 nm) with a resolution of 114,000. They represent the state of the art for the measure of high precision radial velocities: thanks to the simultaneous calibration technique \cite{lovisfischer} and the extreme instrumental stability, ensured by a series of control systems, they can reach a RV accuracy better than 1 m s$^{-1}$. 
In the framework of the GAPS project, HARPS-N spectra are also supported by a coordinated photometric monitoring, in particular for M dwarf stars, performed through the APACHE (Astronomical Observatory of the Autonomous Region of the Aosta Valley, \cite{sozzetti2}) and EXORAP\footnote{See {\it e.g.}, {\tt http://chiantitopics.it/wp-content/uploads/2016/05/Leto.pdf}} (Serra la Nave Observatory) surveys, from the INAF--Astrophysical Observatory of Asiago and by a number of amateur astronomers. 

Thanks to the wide expertise of the GAPS members (high-resolution spectroscopy, stellar activity and pulsations, crowded stellar environments, planetary systems formation, planetary dynamics, data handling) our data are carefully analyzed and discussed within the community, aiming to produce scientific results with a quality as high as possible.
We have developed robust RV data analysis tools (DE-MCMC, Gaussian processes) which are in use to analyze RV data affected by astrophysical noise and to enable the detection of very small amplitude planetary signals. These tools showed their effectiveness also in the international context, as in the "RV challenge" experiment (see \cite{dumusque}) and in the collaboration with the HARPS-N GTO program, allowing a preliminary mass estimate for the first transiting habitable-zone Super-Earth (in a multiple system) around the M0 dwarf K2-3 \cite{crossfield}.

The scientific objectives of GAPS, which are pursued by six sub-programs focused on different type of stars, can be separated into two main aspects:
\begin{itemize}
\item {\bf Planet detection:}
\begin{itemize}
\item determination of the frequency of exoplanets around M stars\footnote{See \cite{perger} for a description of the M dwarf sub-program.}, metal poor stars and stars in open clusters;
\item  search for additional low mass companions in systems that already host giant planets. 
\end{itemize}
\item {\bf System characterization: }
\begin{itemize}
\item Rossiter-McLaughlin effect;
\item Asteroseismology;
\item star-planet interaction;
\item orbital refinement of already known planetary systems.
\end{itemize}
\end{itemize}
In the following Sections we summarize the first results of the GAPS survey.

\section{GAPS Results} 

\subsection{Planet detections}

\begin{itemize}
  \item {\bf XO-2 S b and c} \\
 In 2007 \cite{burke} claimed the presence of a transiting Hot Jupiter around the northern component of the XO-2 binary system, composed by two K0 stars with similar characteristics ($\Delta R \sim0.04$ mag, both super-metal-rich in composition). We performed an intensive monitoring of the southern component with HARPS-N and found evidence of a planetary system composed by a Saturn-mass planet with orbital period of $18.16\pm\rm0.03$ days and a Jupiter-like planet with $P_{orb}=120.80\pm\rm0.34$ days, as presented by \cite{desidera}. This is the first confirmed case of a binary system in which both the components host planets. Moreover, a linear long-term trend is present in the RV residuals of XO-2 S, revealing the possible presence of an additional companion in wide orbit. This system is an interesting laboratory for planet formation and evolution, since the two stellar companions show very similar characteristics but they host two different planetary systems (see Fig. \ref{fig1}). In order to investigate and explain such a configuration, a dedicated study (\cite{damasso1}), including the analysis of the Rossiter-McLaughlin\footnote{see Section \ref{characterization}.} (RML) effect of XO-2 N b, showed that XO-2 N has a larger level of stellar activity with respect to the companion, perhaps due to the presence of the Hot Jupiter, while \cite{biazzo} has shown a significant difference of the chemical abundance between the stars probably correlated with the condensation temperature. A further analysis of this system is ongoing.
\begin{figure}
\centering
 \subfloat
   {\includegraphics[width=.45\textwidth]{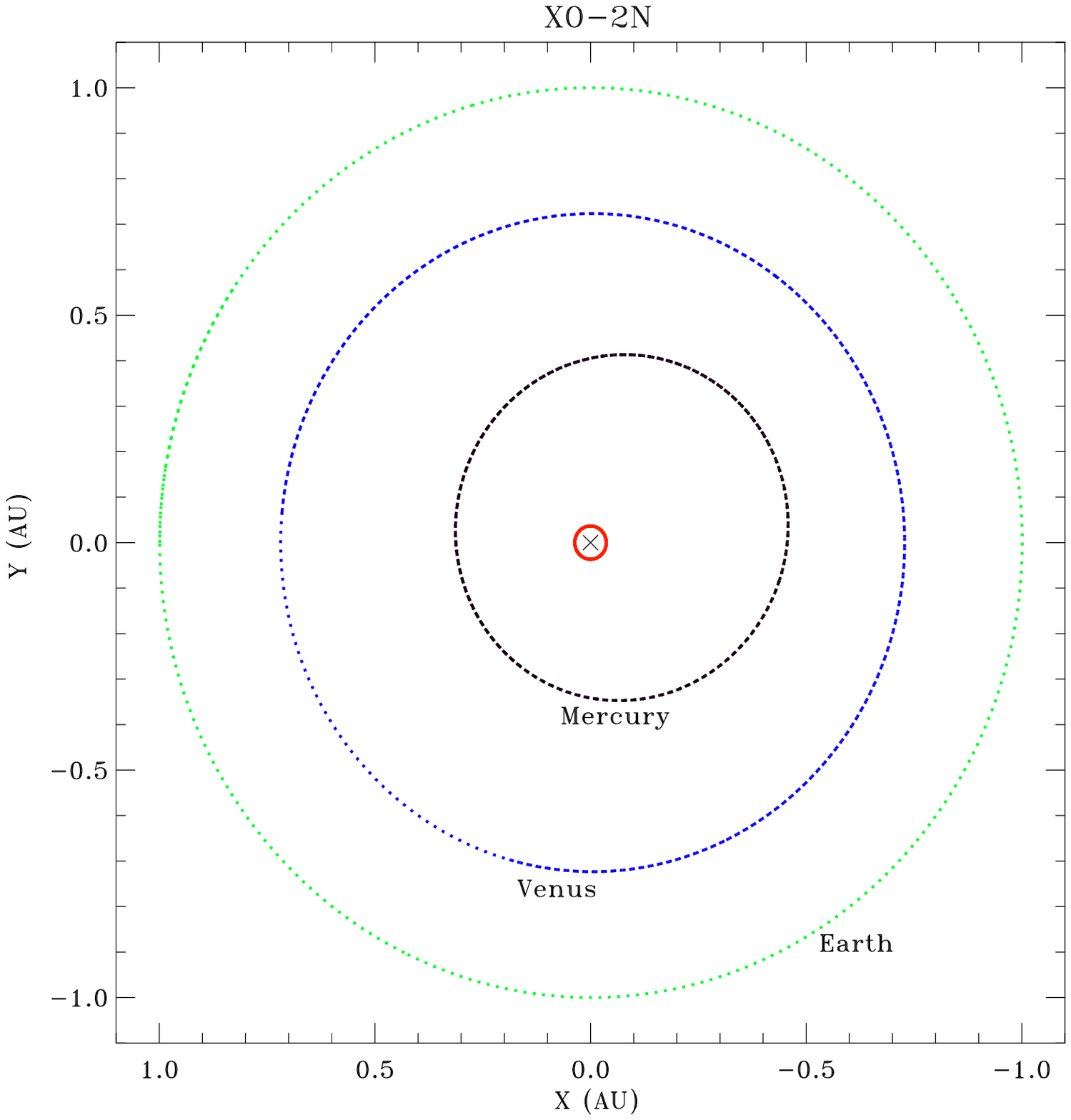}}
 \hspace{5mm}
 \subfloat
   {\includegraphics[width=.45\textwidth]{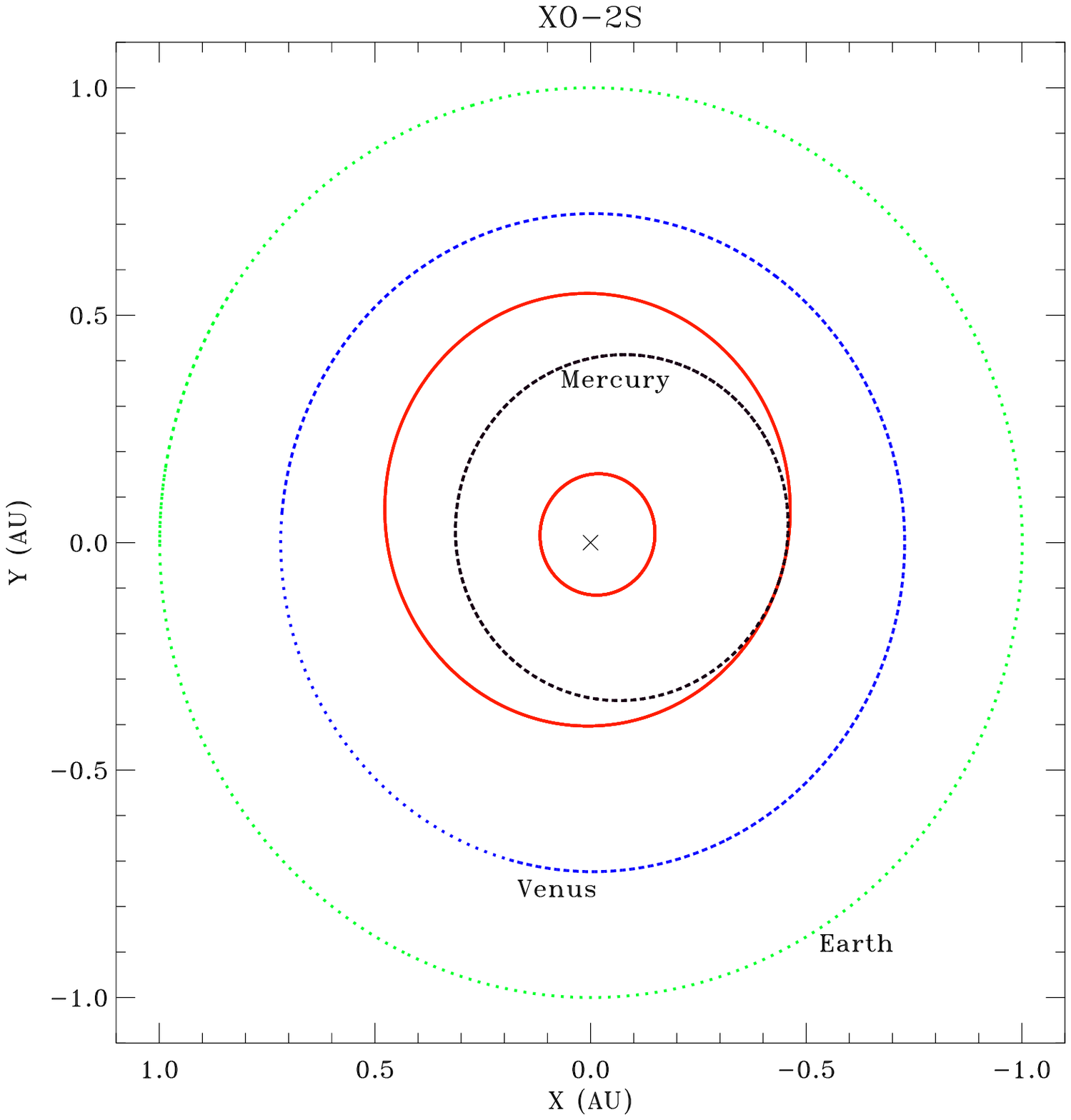}}
\caption{Comparison between the planetary systems orbiting the XO-2 binary stars and the architecture of the inner Solar System. Left panel shows the system of XO-2N, while XO-2S is represented in the right one. The orbits of the XO-2 planets are traced by red ellipses; the orbits of Mercury, Venus and the Earth are indicated with black, blue and green dotted lines respectively. The cross symbol indicates the star location. From \cite{damasso1}.}
\label{fig1}
\end{figure}

 \item {\bf Kelt-6 c} \\
A moderately eccentric giant planet (M$ \sin i=3.7$ M$_{\rm Jup}$) with an orbital period of $1276\pm\rm74$ days has been detected with GAPS observations, as reported by \cite{damasso2}, as an additional companion to the Hot Jupiter, discovered by \cite{collins}, around the F star Kelt-6 (see Figure \ref{fig2}, left panel). 
In \cite{damasso2}, besides the refinement of the orbital solution, the analysis of the RML effect for the inner transiting planet is also presented, Kelt-6 b, showing a slight misalignment of the orbit, with a projected spin-orbit angle, $\lambda $, of $-36\pm\rm11$ degrees. 

 \item {\bf Pr\,0211 c} \\ 
The search for planetary companions around stars in crowded environment is another objective of the GAPS project. \cite{malavolta} report the  RV monitoring with HARPS-N in addition to TRES data available in literature, which allowed to detect a massive planet characterized by high eccentricity ($e$ > 0.60) and long period ($P_{orb}$ > 3500 days) around the target Pr\,0211 in the M44 open cluster, already known to host a Hot Jupiter (\cite{quinn}), see Figure \ref{fig2} (right panel). In this case a deep analysis of the stellar activity has been performed through a dedicated code aiming to exclude a stellar cause for the additional RV signal. This is the first multi-planet system discovered around an open cluster star. Moreover, a significant improvement of the value of the orbital eccentricity ($e = 0.02\pm\rm0.01$) has been determined for the inner planet. 

\begin{figure}
\centering
 \subfloat
   {\includegraphics[width=.5\textwidth]{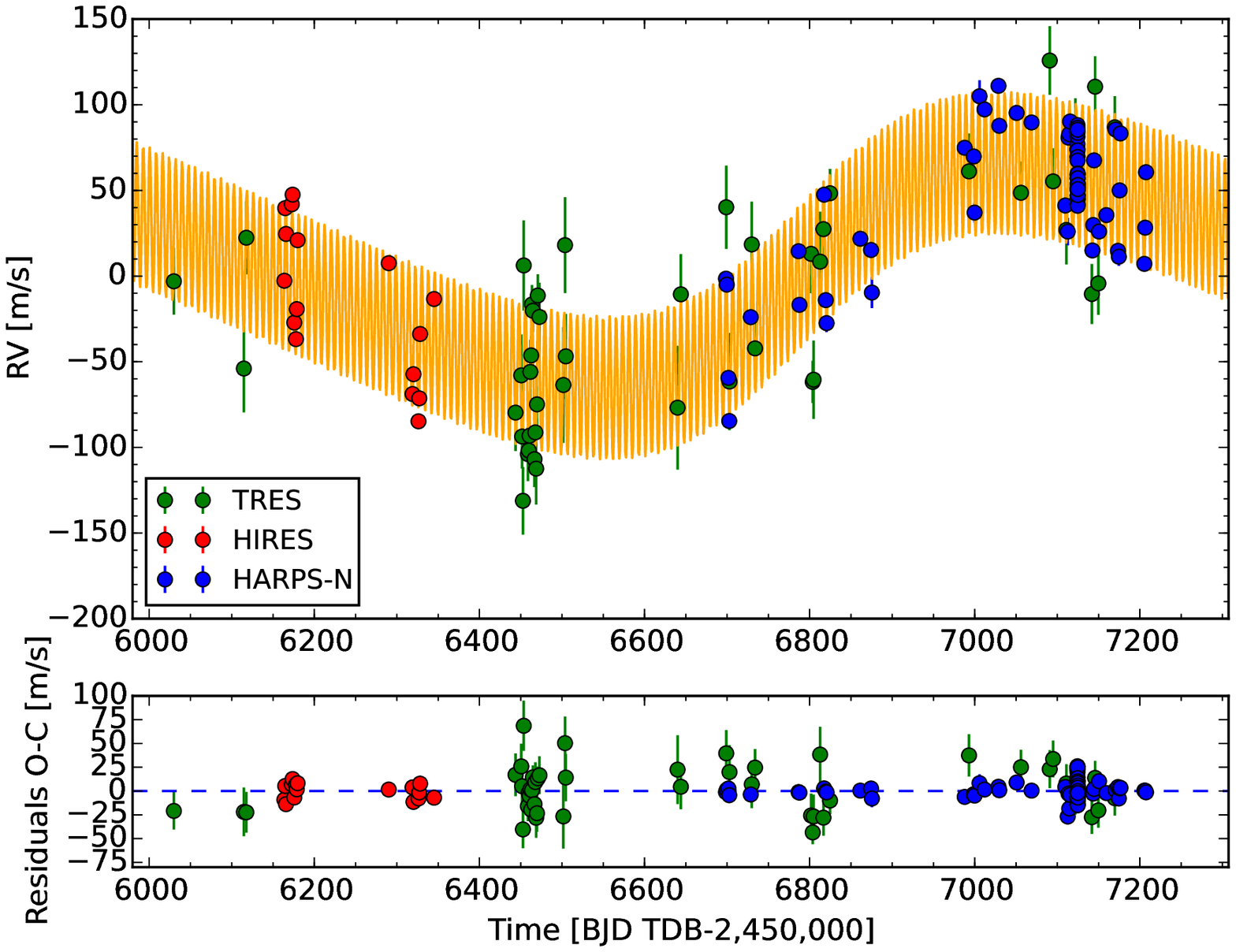}}
 \hspace{5mm}
 \subfloat
   {\includegraphics[width=.4\textwidth]{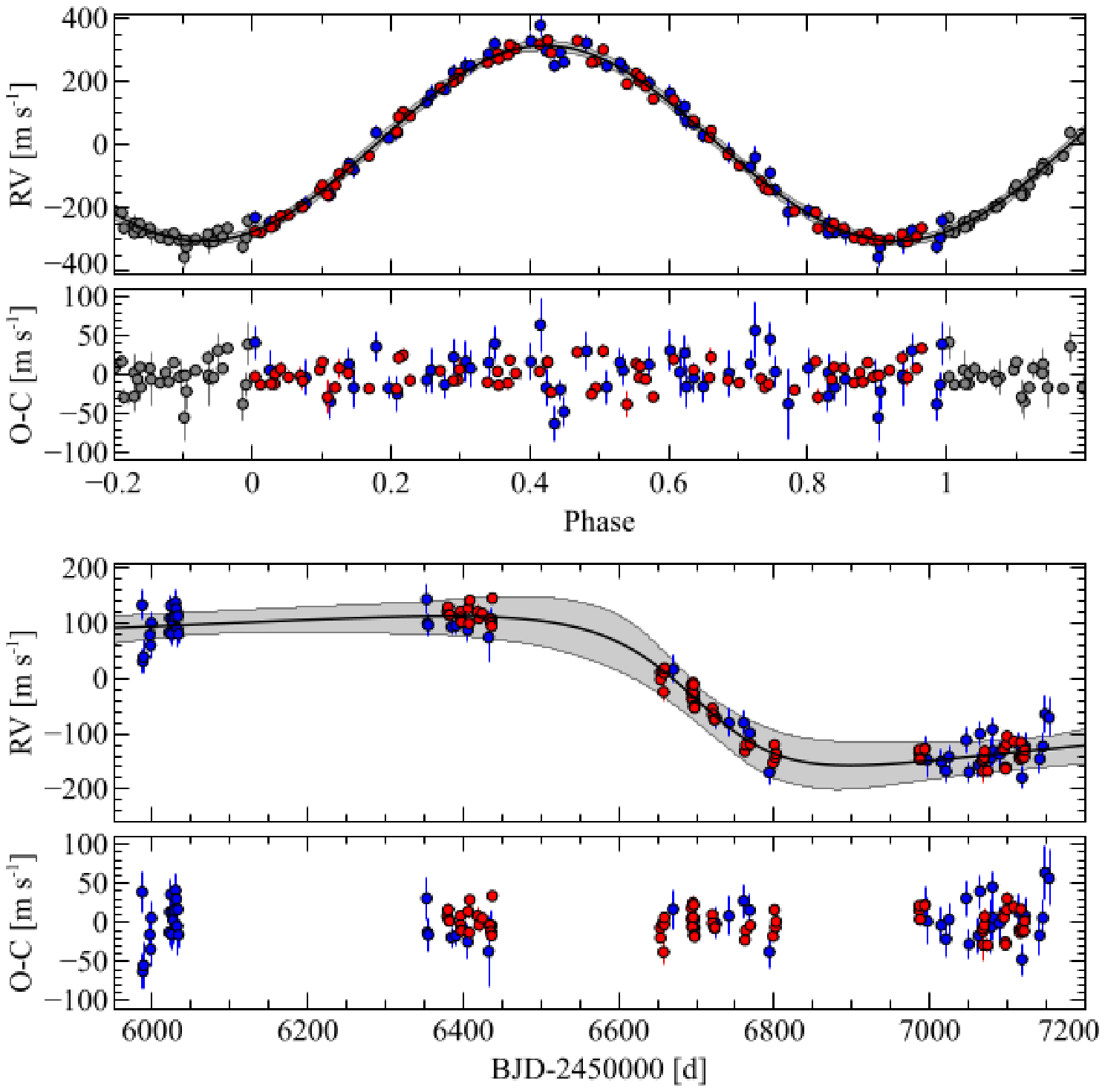}}
\caption{{\it Left panel}: Radial velocity time series of KELT-6 (obtained from HARPS-N, TRES@FLWO-1.5m telescope and HIRES@Keck), over-plotted to the best two-planet Keplerian model (orange line), are shown in the upper panel, while the residuals are in the lower one. From \cite{damasso2}. {\it Right panel}: Orbital solution and RV residuals for Pr\,0211 b (upper panel) \& c (lower panel).  Red and blue dots represent HARPS-N and TRES data, respectively. The gray shaded areas represent the $3 \sigma$ confidence regions. From \cite{malavolta}}
\label{fig2}
\end{figure}

  \item {\bf GJ\,3998 b and c} \\
Within GAPS a parallel and fruitful collaboration has started between the M dwarfs sub-program and some scientists of the ICE (Institut de Ci\`encies de l'Espai/CSIC-IEEC) and the IAC (Instituto de Astrof\'isica de Canarias), called HADES (HArps-n red Dwarf Exoplanet Survey), since they have similar objectives in the study of this type of stars. 
In the framework of this collaboration \cite{affer} presented the detection of a planetary system composed by two Super-Earths around the early M dwarf star GJ\,3998. As for Pr\,0211 the stellar activity signal can be confused with the RV modulation induced by the presence of low-mass planets, so the analysis was carried on in order to properly treat this phenomenon. As a result, signatures of both the stellar and the differential rotation were identified in the periodogram of the RV, in addition to the ones of the two planets (Figure \ref{fig3}). The inner planet shows an orbital period of $2.6498\pm\rm0.0008$ days and a minimum mass of $2.47\pm\rm0.27$ M$_{\oplus}$, while the outer shows a $13.74\pm\rm0.02$ days period and a minimum mass of $6.26\pm\rm0.77$  M$_{\oplus}$.
\end{itemize}
\begin{figure}
\centering
\includegraphics[width=.6\textwidth]{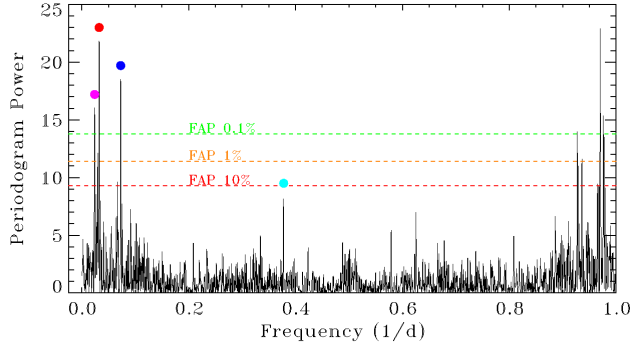}
\caption{GLS periodograms of the radial velocities of GJ 3998: the periodicities of the two planets are marked with a blue (13.74 days) and a cyan dot (2.65 days), the stellar (30.7 days) and the differential rotation (42.5 days) are indicated by red and magenta dots, respectively. The dashed lines indicate 0.1\%, 1\% and 10\% level of false alarm probability. From \cite{affer}}
\label{fig3}
\end{figure}

\subsection{System characterization} \label{characterization}
\begin{itemize}
  \item {\bf Rossiter-McLaughlin effect} \\
The Rossiter-McLaughlin (RML) effect occurs when a planet transits in front of the disk of its host star, producing an anomaly in the radial velocities. Due to the temporary blocking of a fraction of the blue-shifted region or a fraction of the red-shifted region of the star, the resulting spectrum appears to be slightly red-shifted and blue-shifted respectively.
This effect is useful to characterize the orbit of the planet companion, allowing to measure the sky projected angle $\lambda$ between the star spin axis and the planet orbital axis. With a dedicated program of GAPS we observed the RML effect for several systems, showing {\it e.g.} the retrograde orbit of HAT-P-18 b \cite{esposito} and the good degree of alignment for Qatar-1 b \cite{covino} and for HAT-P-36 and WASP-11/HAT-P-10 \cite{mancini} (see Figure \ref{fig4}, left panel). The very last result of the RML sub-program reports the orbital obliquity of the three very massive close-in giant planets around WASP-43, HAT-P-20 and Qatar-2 \cite{esposito2}.

  \item {\bf Asteroseismology and star-planet interaction } \\
 The synergy between Asteroseismology and the exoplanets search is acknowledged and powerful, since the planet parameters are strictly correlated with the stellar ones, available today with higher accuracy thanks to the seismic analysis of stellar oscillations. A pilot study was performed within GAPS, aiming to search for stellar oscillations in $\tau$ Boo. In \cite{borsa} the frequency of the maximum power $\nu _{\rm max}$ and the large separation $\Delta \nu$, two quantities which are directly connected with the mean properties of the star (see \pos{Di Mauro} for a review), are measured. This result encourages the application of the asteroseismic approach to stars hosting exoplanets, in order to constrain their ages and masses. Finally, a full analysis of the activity indicators have been performed in order to detect the possible star-planet interaction, which still remain ambiguous for $\tau$ Boo. 

In \cite{maggio} a clear indication of star-planet interaction between the G-star HD\,17156 and the hosted Hot Jupiter has been found, thanks to a coordinated HARPS-N and XMM observations. A significant X-ray detection and a strong increasing of the Ca II H\&K chromospheric index occurred near the periastron visit of this planet, characterized by a high eccentricity.

  \item {\bf The curious case of TrES-4 b} \\
Within GAPS we performed a RV monitoring to refine the orbital solution of the transiting Hot Jupiter TrES-4 b (\cite{mandushev}), with a further support of photometric data. \cite{sozzetti} determined a value of the RV semiamplitude compatible with a planetary mass significantly lower than previously reported ($0.49\pm\rm0.04$ M$_{\rm Jup}$), making TrES-4 b the transiting Hot Jupiter with the second-lowest density known (see Figure \ref{fig4}, right panel). This result opens several questions about the formation of such highly inflated bodies.
\begin{figure}
\centering
 \subfloat
 {\includegraphics[width=.45\textwidth]{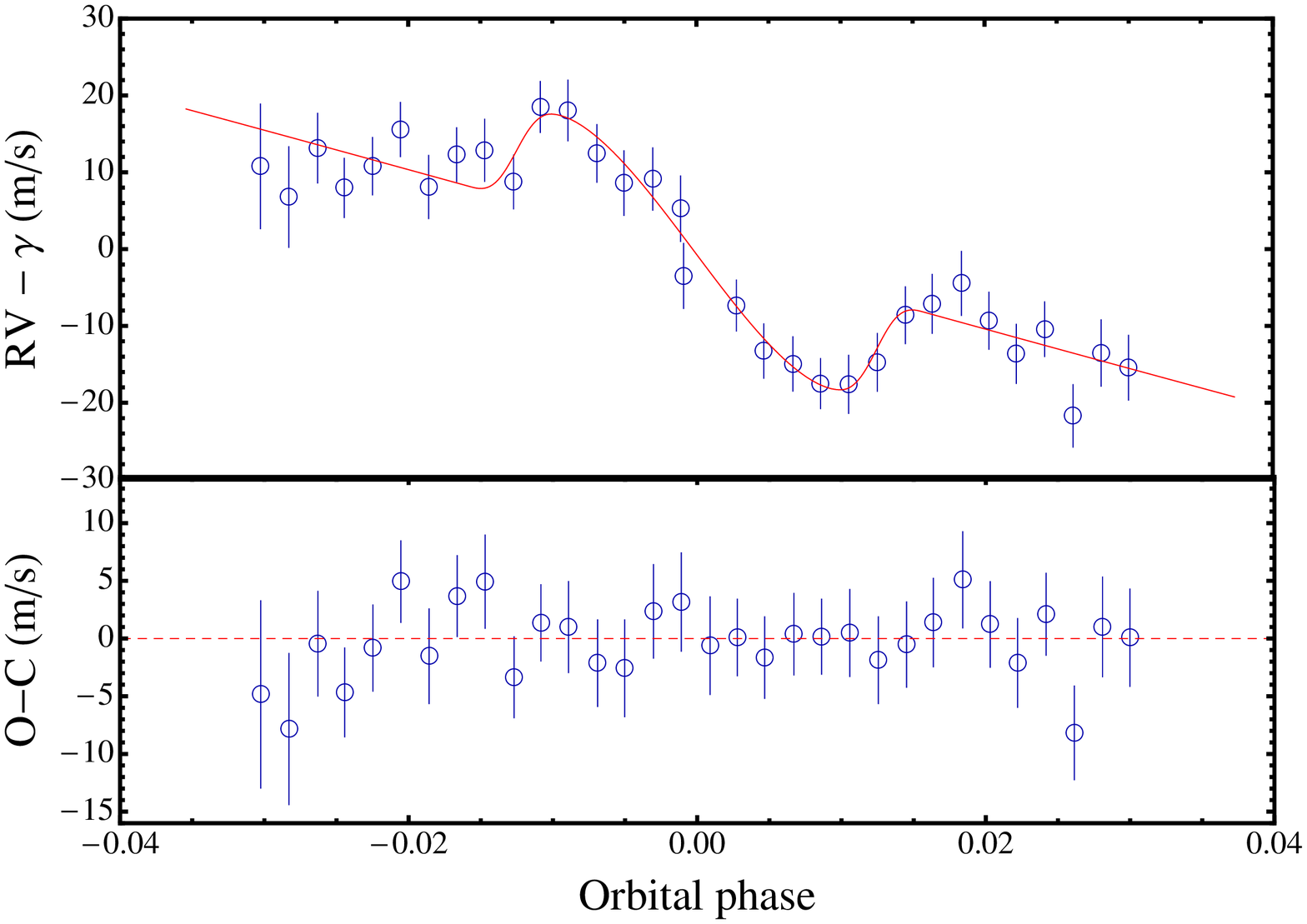}}
 \hspace{5mm}
 \subfloat
   {\includegraphics[width=.4\textwidth]{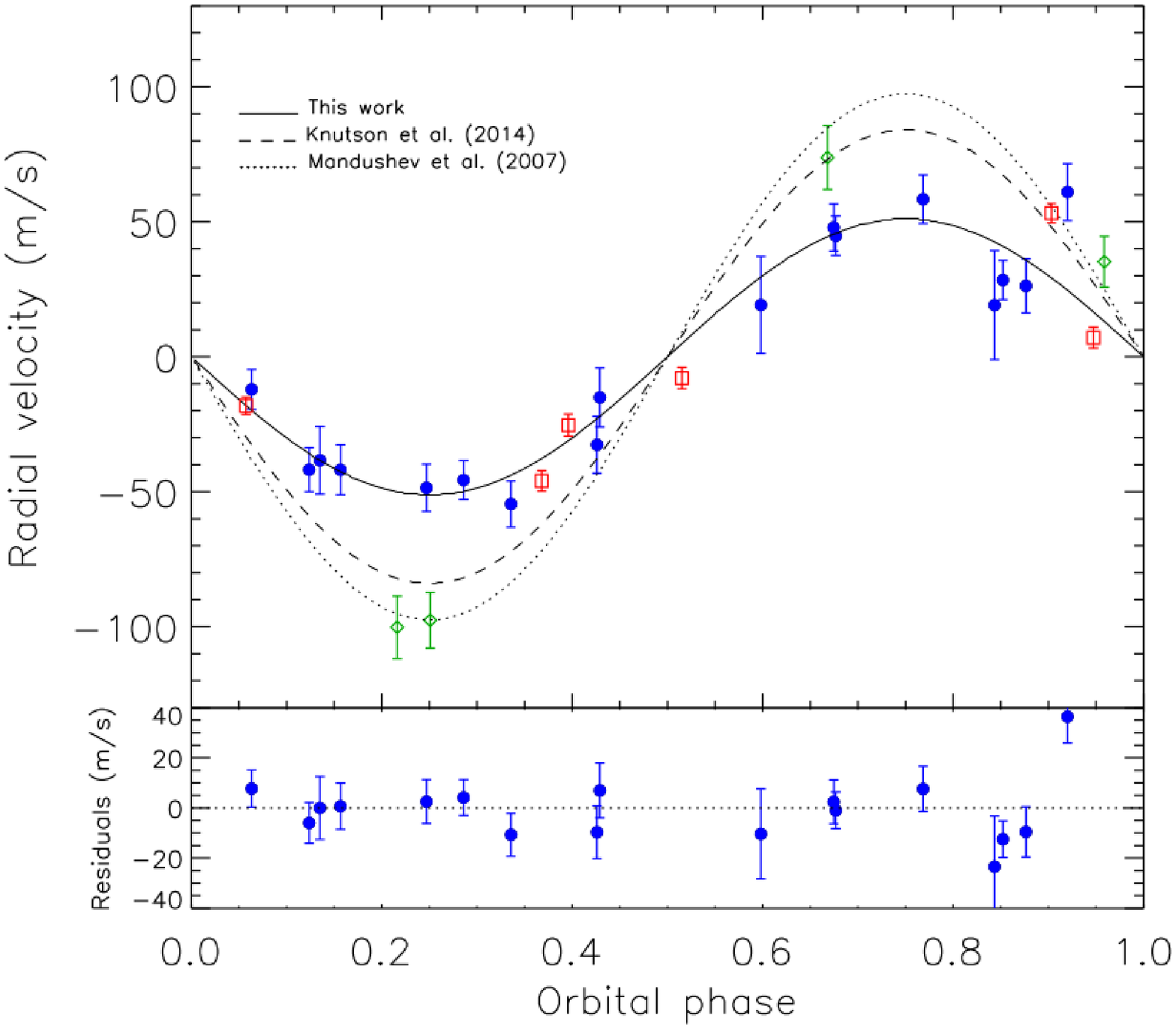}}
\caption{{\it Left panel}: Phase-folded RV data of the Rossiter-McLaughlin effect of WASP-11 b observed with HARPS-N over-plotted to the best-fitting model. The corresponding residuals are plotted in the lower panel. From \cite{mancini}. {\it Right panel}: Phase-folded RV measurements of TrES-4 obtained with HARPS-N (blue circles) and the best-fit Keplerian orbit model (black solid line). The comparison with previous studies clearly shows a discrepancy of the RV semiamplitude (\cite{mandushev}, green diamonds and \cite{knutson}, red squares). The lower panel shows the residuals from the best-fit model to the HARPS-N radial velocities. From \cite{sozzetti}}
\label{fig4}
\end{figure}

HARPS-N spectra also allowed to confirm or rule-out previous planet detections, as in the case of the metal poor star HIP\,11952. In \cite{desidera2} we show that our radial velocities scatter is incompatible with the presence of the two giant planets claimed by \cite{setiawan}, whose data were affected by the lower sensibility of the instrumentation.
This result dismissed planets around a very metal poor star, generally excluded by the planet formation theory.

  \item {\bf Stellar parameters and activity for M stars} \\
The HARPS-N spectra used to obtain high precision radial velocities were also exploited in a parallel study by \cite{maldonado} to obtain 
temperature-sensitive ratios of spectral features from a large sample of M0-M4.5 dwarf stars. They provide empirical calibrations for masses, radii, and gravities as a function of effective temperature and metallicity. Particular attention is also paid on the characteristics of the stellar activity of M dwarfs: \cite{maldonado2} presented a study on the relationship between activity-rotation and stellar parameters and the flux-flux relationship of different chromospheric lines, while \cite{scandariato} investigated the chromospheric emission of Ca II H\&K and H${\alpha}$ for low-activity early-M dwarfs, with the contribution of the HADES collaborators.

\end{itemize}

\subsection{Further works} 
Other studies involved in particular the system characterization, {\it e.g.}, the Bayesian homogeneous determination of orbital and physical parameters for a large sample of giant transiting planets (\cite{bonomo}), and the full analysis of the HD\,108874 planetary system, reported in \cite{benatti}. New planet detections will be soon presented in dedicated papers, while the identified planet candidates still require data for a robust confirmation.

\section{Conclusions and future perspectives}
GAPS has coagulated a large number of Italian scientists involved in the exoplanets field, resulting in a strong and coordinated collaboration.
Several interesting results were published up to now and even more papers are in preparation or close to the submission. A number of planet candidates has been identified but in some cases we need more data for their confirmation, so we plan further observations. After four years of HARPS-N observations and analysis we have developed an optimized observing strategy and new analysis tools, in particular for those objects which require many data and specific treatment of the RV time series.
Anyway, new perspectives are foreseen for GAPS, since GIARPS is expected to be available at TNG  with high impact on the exoplanetary research. In fact GIARPS will combine HARPS-N and GIANO, the high resolution spectrograph of TNG in the near infrared (see \cite{origlia}), providing simultaneous observations in the visible and in the near infrared bands, see \pos{Claudi} (this proceeding) for details.  
The extension of the wavelength range will open to the GAPS community new scenarios and objectives in the study of the extrasolar planets.

\end{document}